\documentclass[npg,2-column]{copernicus}

\usepackage{amssymb}

\newcommand{\ignore}[1]{}
\newcommand{\gn}{\mathbf g}
\newcommand{\un}{\mathbf u}
\newcommand{\vn}{\mathbf v}
\newcommand{\rn}{\mathbf r}
\newcommand{\xn}{\mathbf x}

\newcommand{\Omegan}{\mathbf \Omega}
\newcommand{\Wn}{\mathbf W}

\begin{document}
\nolinenumbers

\title{Modeling the dynamical sinking of biogenic particles in oceanic flow}


\Author[1]{Pedro}{Monroy}
\Author[1]{Emilio}{Hern\'{a}ndez-Garc\'{\i}a}
\Author[1,2]{Vincent}{Rossi}
\Author[1]{Crist\'{o}bal}{L\'{o}pez}

\affil[1]{IFISC, Instituto de F\'\i sica Interdisciplinar y
Sistemas Complejos (CSIC-UIB), 07122 Palma de Mallorca, Spain}
\affil[2]{Mediterranean Institute of Oceanography (UM 110, UMR 7294), CNRS, Aix Marseille Univ., Univ. Toulon, IRD, 13288, Marseille, France}



\runningtitle{Sinking of biogenic particles in oceanic flow}

\runningauthor{Monroy et al.}

\correspondence{Pedro Monroy (pmonroy@ifisc.uib-csic.es)}

\received{}
\pubdiscuss{} 
\revised{}
\accepted{}
\published{}


\firstpage{1}

\maketitle

\begin{abstract}
  We study the problem of sinking particles in a realistic oceanic flow, with
  major energetic structures in the mesoscale, focussing in the range of
  particle sizes and densities appropriate for marine biogenic particles. Our
  aim is to evaluate the relevance of theoretical results of finite size
  particle dynamics in their applications in the oceanographic context. By
  using a simplified equation of motion of small particles in a mesoscale
  simulation of the oceanic velocity field, we estimate the influence of
  physical processes such as the Coriolis force and the inertia of the
  particles, and we conclude that they represent negligible corrections to the
  most important terms, which are passive motion with the velocity of the
  flow, and a constant added vertical velocity due to gravity. Even if within
  this approximation three-dimensional clustering of particles can not occur,
  two-dimensional cuts or projections of the evolving three-dimensional
  density can display inhomogeneities similar to the ones observed in sinking
  ocean particles.
\end{abstract}

\introduction  

The sinking of small particles suspended in fluids is a topic of both
fundamental importance and of practical implications in diverse fields ranging
from rain nucleation to industrial processes
\citep{Michaelides1997,Falkovich2002}.

In the oceans, photosynthesis by phytoplankton in surface waters uses
sunlight, inorganic nutrients and carbon dioxide to produce organic matter
which is then exported downward and isolated from the atmosphere
\citep{Henson2012}, a process which forms the so-called biological carbon
pump. The downward flux of carbon-rich biogenic particles from the marine
surface due to gravitational settling, one of the key process of the
biological carbon pump, is responsible (together with the solubility and the
physical carbon pumps) of much of the oceans' role in the Earth carbon cycle
\citep{Sabine2004}. Although most of the organic matter is metabolized and
remineralized in surface waters, a significant portion sinks into deeper
horizons.  It can be sequestered on various time scales spanning a few years
to decades in central and intermediate waters, several centuries in deep
waters and up to millions of years locked up in bottom sediments
\citep{Devries2012}. Suitable modeling of the sinking process of particulate
matter is thus required to properly assess the amount of carbon sequestered in
the ocean and in general to better understand global biogeochemical cycling
and its influence on the Earth climate.


This is a challenging task that involves the downward transport of particles
of many different sizes and densities by turbulent ocean flows which contain
an enormous range of interacting scales. In the oceanographic community,
numerous studies approached this problem by considering biogenic particles
transported in oceanic flow as passive particles with an added constant
velocity in the vertical to account for the sinking dynamics
\citep{Siegel1997,Siegel2008,Qiu2014,Roullier2014,Vansebille2015}.  They
suggest that the sinking of particles may not be strictly vertical but
oblique, meaning that the locations where the particles are formed at the
surface may be distant from the location of their deposition in the seafloor
sediment. Then \cite{Siegel2008} presented the concept of statistical funnels
which describe and quantify the source region of a sediment trap (subsurface
collecting device of sinking-particles used to get estimate of vertical
fluxes).  The validity of this approximation and the influence of different
physical processes is however poorly discussed in these analyses.

In the physical community, the framework to model sinking particles is based
on the Maxey-Riley-Gatignol equation for a small spherical particle moving in
an ambient flow
\citep{Maxey1983,Gatignol1983,Michaelides1997,Provenzale1999,Cartwright2010},
which highlights the importance of mechanisms beyond passive transport and
constant sinking velocity, such as the role of finite size, inertia and
history dependence.  A major outcome of these studies is that inhomogeneities
and particle clustering can arise spontaneously even if the fluid velocity
field is incompressible and particles do not interact \citep{Squires91}.
Particle clustering and patchiness is indeed observed in the surface and
subsurface of the ocean \citep{Logan1990,Buesseler2007,Mitchell2008}

Here we consider the theory of small but finite-size particles driven by
geophysical flows, which is, as mentioned above, conveniently based on the
Maxey-Riley-Gatignol equation. In Sect. \ref{sec:propertiesmarineparticle} we
review the main characteristics of marine particles which are relevant for
their sinking dynamics. In Sect. \ref{sec:equations} we present the equations
of motion describing this process, together with the approximations required
to obtain them and the type of particles for which they are valid. In
particular, we discuss its validity and the relevance of the different
physical processes involved in the range of sizes and densities of marine
biogenic particles. In Sect. \ref{sec:simulations} we use these equations to
study the settling dynamics in a modelled oceanic velocity field produced by a
realistic high-resolution regional simulation of the Benguela upwelling system
(southwest Africa). We estimate the relevance of physical processes such as
the Coriolis force and the inertia of the particles with respect to the
settling velocity. We also observe the spatial distribution of particles
falling onto a plane of constant depth above the seabed and we identify
clustering of particles that is interpreted with simple geometrical arguments
which do not require physical phenomena beyond passive transport and constant
terminal velocity. Our main results are finally summarized in a Conclusion
section.



\section{Characteristics of marine biogenic particles}
\label{sec:propertiesmarineparticle}

In theory, the sinking velocities of biogenic particles depend on various
intrinsic factors (such as their sizes, shapes, densities, porosities) which
can be modified along their fall by complex bio-physical processes
(e.g. aggregation, ballasting, trimming by remineralisation) as well as by the
three-dimensional flow field \citep{Stemmann2012}. However reasonable
estimates of the effective sinking velocities of marine particles can be
obtained by taking into account only its size and density
\citep{McDonnell2010}.  In our Lagrangian setting we thus consider that the
two key properties of marine particles controlling their sinking dynamics are
their size and density.  Here we present the standard classification of marine
particles according to the typical range of size and density by compiling
different bibliographical sources.

\subsection{Size}

Because of the diversity of the shapes, the size of a particle refers to the
diameter of a sphere of equivalent volume (Equivalent Spherical Diameter)
\citep{Guidi2008}. The size of marine particles ranges from $1~nm$
(almost-dissolved colloids) to aggregates larger than $1~cm$
\citep{Stemmann2012}.

\begin{figure}[t]
  \includegraphics[width=\columnwidth]{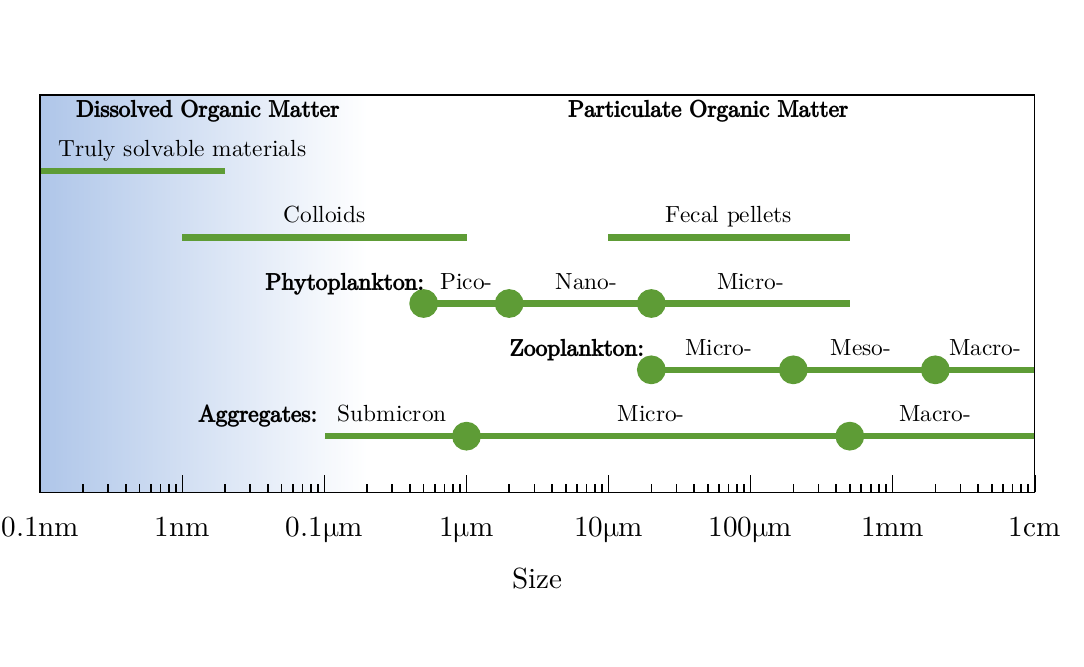}
  \caption{Size and classification of marine particles (adapted from
    \citet{Simon2002}).  }
  \label{fig:sizeparticles}
\end{figure}

Originally, the size classification of particles was based on the minimal pore
size of the nets used for their collection, which is about
$\simeq 0.45 - 1.0~\mu m$. Any material larger than $0.2~\mu m$ (thus isolated
by the filtration of seawater) is regarded as particulate organic matter,
while the fraction that percolates through the filter is labelled as dissolved
matter. This includes colloidal and truly dissolved materials (see
Fig. \ref{fig:sizeparticles}). Although this discrimination of the
size-continuum observed in the real ocean is somehow arbitrary, it is useful
--and we will follow it-- because particles smaller than $1.0~\mu m$ are not
prone to sinking \citep{Hedges2002}.

In the following, our focus is thus on particulate matter larger than
$1.0~\mu m$ (Fig. \ref{fig:sizeparticles}). Organic matter is produced in the
sun-lit layer of the ocean by the primary production through photosynthesis of
autotrophic microbes (mainly bacteria and phytoplankton). During their
lifetime growth they exude colloidal and small particles to finally form
larger particles when they die. Dead phytoplankton are within the range of
$1~\mu m$ (picoplankton, e.g.  cyanobacteria) and a few hundred of micrometers
(microphytoplankton, e.g.  diatoms).

Thereafter zooplankton consumes alive phytoplankton and inert particles and
produce fecal pellets and dead bodies. Most fecal materials have enough size
to sink rapidly by their own \citep{DeLaRocha2007}.  Typical sizes of such
particles are $10~\mu m$ for a pellet of copepod of $200~\mu m$ length
\citep{Jackson2001}, krill fecal pellets are between $160~\mu m - 460~\mu m$
\citep{McDonnell2010} and euphausiid fecal pellets span $300~\mu m - 3~mm$
\citep{Komar1981}, providing the total range of $10 \mu m$ to
$3~mm$. Concerning the zooplankton dead bodies, they are divided in micro-,
meso- and macro-, with sizes in the range $20 \mu m -1 cm$. A detailed summary
is given in Table \ref{tab:marineparticlesizes}.


Finally, there are the so-called organic aggregates which occur in the size
range of $1\mu m$ to $10cm$. They are typically formed in situ by
physical aggregation or biological coagulation and are usually composed of
numerous planktonic individuals and fecal pellets sticked together within a
colloidal matrice. They are often distinguished in three size classes
\citep{Simon2002}: macroscopic aggregates or macro-aggregates $>5 mm$ usually
called marine snow; microscopic, from $1$ to $500\mu m$, also known as
micro-aggregates; and submicron particles $<1\mu m$ (which do not sink).

\begin{table*}[t] \centering \footnotesize
  \begin{tabular}{|p{9cm}|p{4,5cm}|}\hline Individual Particles (mostly
organic) & Aggregates (compounds of organic and inorganic particles) \\ \hline
Fecal pellets (cylindrical):
    \begin{itemize}
    \item Krill fecal pellets: Length between $400~\mu m$ and $9~mm$, diameter
$120~\mu m$ \citep{McDonnell2010}. ESD ($160~\mu m - 460~\mu m$)
    \item $10~\mu m$, consistent with pellet volume of a $200~\mu m$ copepod
\citep{Jackson2001}
    \end{itemize}

    Dead zooplankton \citep{Stemmann2012}:
    \begin{itemize}
    \item Macrozooplankton:\newline$\text{size}>2000~\mu m$
    \item Mesozooplankton:\newline$200<\text{size}<2000~\mu m$
    \item Microzooplankton:\newline$20<\text{size}<200~\mu m$
    \end{itemize}

    Dead phytoplankton \citep{Stemmann2012}:
    \begin{itemize}
    \item Microphytoplankton:\newline(size$>200~\mu m$)
    \item Nanophytoplankton:\newline($20<$size$<200~\mu m$)
    \item Picophytoplankton:\newline($2<\text{size}<20~\mu m$)

    \end{itemize} & Aggregates\citep{Simon2002}:
                    \begin{itemize}
                    \item Macroscopic (Marine Snow):\newline
                      $\text{size}>500~\mu m$.
                    \item Microscopic:\newline
                      $1\mu m<\text{size}<500~\mu m$.
                    \item Submicron:\newline
                      $\text{size}<1~\mu m$.
                    \end{itemize}\\ \hline
  \end{tabular} \\
  \caption{Simplified categorization of marine biogenic particles, and their
    associated sizes.}
  \label{tab:marineparticlesizes}
\end{table*}

\subsection{Density}
The density of marine particles depends on their composition
which can be divided into a mineral and  a organic fraction
\citep{Maggi2015}. The mineral or inorganic matter consists of
biogenic minerals: Particulate Inorganic Carbon (PIC), e.g.
calcium carbonate produced by coccoliths with density
$2700~kg/m^3$ and Biogenic Silica (BSi), produced by diatoms,
significantly less denser than PIC, $1950~kg/m^3$
\citep{Balch2010}. The density of Particulate Organic Matter
(POC) ranges widely depending on its origin. For instance, the
density of cytoplasm spans from $1030$ to $1100~kg/m^3$, while
the one of fecal pellets ranges between $1174~kg/m^3$ and
$1230~kg/m^3$ \citep{Komar1981}. Despite this variability, it
is possible to assign a range to the density of organic matter,
from $1050$ to $1500~kg/m^3$.

Considering all these estimates together, the density of marine particle
ranges approximately between $1050$ to $2700~kg/m^3$ \citep{Maggi2013}. This
should be compared to standard values for sea water density in the interior
ocean which spans roughly 1020-1030 $kg/m^3$. Thus most of the particle types
described previously will sink. Assuming constant size and density for each
particle along its downward course, we deduce that most of the particles types
described previously will sink. This holds without considering biogeochemical
and (dis)aggregation processes that may occur in nature, thus lowering the
particle density and resulting in clustering and trapping of particles at
particular isopycnals \citep{Sozza2016}. Note that we do not consider here
living organisms which show vertical movements by active swimming or by
controlling their buoyancy \citep{Moore1996,Azetsu2004}.

\section{Equations of motion for small spherical rigid particles}
\label{sec:equations}

\subsection{The Maxey-Riley-Gatignol equation}

To describe the sedimentation of biogenic particles, we need to
study the motion of single particles driven by fluid flow. A
milestone to analyze the dynamics of a small spherical rigid
particle of radius $a$ subject to gravity acceleration $\gn$ in
an unsteady fluid flow $\un (\rn,t)$ is given by the
Maxey-Riley-Gatignol \citep{Maxey1983,Gatignol1983,Michaelides1997,Cartwright2010} equation
(MRG in the following):
\begin{equation}\label{eq:MRG}
\begin{split}
    \rho_p \frac{d\vn}{dt} =& \rho_f \frac{D\un}{Dt} + (\rho_p-\rho_f)\gn
    - \frac{9 \nu \rho_f }{2 a^2}\left(\vn - \un - \frac{a^2}{6}\nabla^2 \un\right) \\
    &-\rho_f \left(\frac{d\vn}{dt}-\frac{D}{Dt}(\un + \frac{a^2}{10}\nabla^2 \un)\right)\\
    &  -\frac{9 \rho_f}{2 a}\sqrt{\frac{\nu}{\pi}} \int_0^t \frac{\frac{d}{ds}(\vn - \un - \frac{a^2}{6} \nabla^2 \un)}{\sqrt{t-s}}ds  .
\end{split}
\end{equation}
The velocity of the particle is denoted by $\vn=\vn(t)$. The
particle and fluid densities are $\rho_p$ and $\rho_f$,
respectively, and $\nu$ denotes the fluid kinematic viscosity.
The time derivative operators $\frac{d}{dt} = \frac{\partial
}{\partial t} + \vn \cdot \nabla $ and $\frac{D}{Dt} =
\frac{\partial }{\partial t} + \un \cdot \nabla $ denote the
time rate of change following the particle itself and the time
rate of change following a fluid element in the undisturbed
flow field $\un(\rn,t)$ respectively. This equation of motion
gives the balance between the different forces acting on the
particle, which correspond to the right-hand-side terms: the pressure
force (the force exerted on the particle by the undisturbed
flow), the buoyancy force, the drag force (Stokes drag), the
added mass force resulting from the part of the fluid moving
with the particle, and the history force. As will be discussed
below the validity of this equation requires several
conditions, being the main one the small size of the particles.
The terms with $a^2 \nabla^2 \un$ are the Fax\'{e}n corrections
\citep{Faxen1922}.

The full MRG is very complicated to manage. A further simplification is
usually performed based on the single assumption of very small particles (what
this exactly means will be discussed later on). With this, the Fax\'{e}n
corrections and, as commented below, also the history term (since
$a/\sqrt{\nu} << 1$) can be neglected
\citep{Maxey1983,Michaelides1997,Haller2008}. Note however that the history
term can be relevant under some conditions, as for example larger particle
size \citep{Daitche2011,Guseva2013, Guseva2016, Olivieri2014}.  Thus we obtain
the standard form of the MRG equations \citep{Maxey1983}:
\begin{equation}
\frac{d\vn}{dt}=\beta \frac{D\un}{Dt} + \frac{\un - \vn + \vn_s}{\tau_p},
\label{eq:MRsimp}
\end{equation}
where $\beta=\frac{3\rho_f}{2\rho_p+\rho_f}$, the Stokes time is
$\tau_p=\frac{a^2}{3\beta \nu}$, and $\vn_s=(1-\beta) \gn \tau_p$ is the
settling velocity in quiescent fluid.
Equation (\ref{eq:MRsimp}) is the starting point for most
inertial particle studies \citep{Michaelides1997,Balkovsky2001,Cartwright2010}.

We now discuss the validity of the MRG equation Eq.  (\ref{eq:MRG}) or rather
its simplified form Eq.  (\ref{eq:MRsimp}) for the range of sizes and
densities of marine organisms.  We do so in the context of open-ocean flows,
which are typically most energetic at the mesoscale (scales of about 100 km),
and where there is a strong stratification, with vertical velocities three or
four orders of magnitude smaller than horizontal ones. The motion becomes more
three-dimensional, and then the concepts of three-dimensional turbulence more
relevant, below scales $l$ of some hundred of meters, with typical velocities
decreasing as $l^{1/3}$ for decreasing scale and velocity gradients increasing
as $l^{-2/3}$ until the Kolmogorov scale $l=\eta$ below which flow becomes
smooth. Because of its direct exposure to wind, turbulence intensity is
typically larger at the ocean surface, with values of turbulent energy
dissipation in the range
$1\cdot 10^{-6}m^2/s^3<\epsilon<3\cdot 10^{-5} m^2/s^3$ \citep{Jimenez1997},
than at depth. The first condition for the validity of the MRG equation that
was originally discussed by \cite{Maxey1983} is that the particles have to be
much smaller than the typical length scale of variation of the flow.  This
means that for multiscale (turbulent) flows the radius of the particle $a$ has
to be much smaller than the Kolmogorov scale $\eta$, which according to the
previous values of $\epsilon$, is typically $0.3 mm < \eta < 2 mm$ in the
ocean surface \citep{Okubo1971,Jimenez1997}.  Note that we only have to
consider worst-case situations for assessing the validity of the different
approximations.  Another condition to be fulfilled is that the shear Reynolds
number must be small $Re_{\nabla}=a^2U/\nu L<<1$, where $U$ and $L$ are
typical velocity and length scales. For a turbulent ocean with multiple scales
and velocities, the most restrictive condition arises when they take the
values of the Kolmogorov velocity $v_{\eta}$ and length $\eta$, respectively,
since then the velocity gradients are maxima.  In this case the condition
becomes $Re_{\nabla}=a^2/\eta^2<<1$, which again is satisfied for small
particles.  We note that \citet{Guseva2013} found that the relative importance
of the history term in Eq.  (\ref{eq:MRG}) with respect to the drag force is
of the order of a parameter which in our notation is $(Re_\nabla)^{1/2}$.
This justifies neglecting the history term for small particles, although its
importance increases for increasing size \citep{Daitche2011, Guseva2013}.

\begin{figure}[t]
\includegraphics[width=\columnwidth]{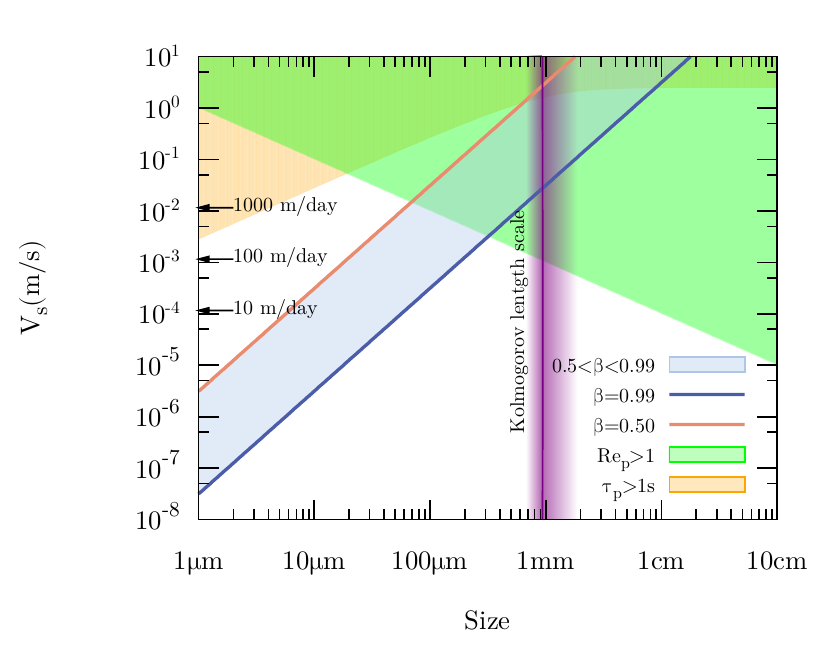}
\caption{Sinking velocity versus particle radius for different $\beta$, which
  is determined by densities.  The blue zone determines the values of the
  settling velocities at a given radius, as determined by the typical marine
  particle densities. The green area is determined by the condition $Re_p >1$
  for which the MRG equation is not valid. Use of the MRG equation is also
  unjustified for particles larger than the Kolmogorov length scale also
  plotted in the figure.  We also show the region
  $\tau_p>\tau_\eta\approx 1 s$ where the additional approximation leading to
  Eq. (\ref{eq:final}) becomes invalid.}
\label{fig:modelvalidity}
\end{figure}

Another condition to be satisfied for the validity of the MRG equation is that
the so-called Reynolds particle number, $Re_p =\frac{a |\vn - \un |}{\nu}$
should fulfill $Re_p <<1$.  Considering that gravity force dominates over
other forces one has $|\vn-\un|\simeq |\vn_s|\equiv v_s$, where $\vn_s$ is, as
introduced before, the settling velocity of particles in a quiescent fluid due
to Stokes drag. The Reynolds particle number is then $Re_p=\frac{a
  v_s}{\nu}$. Note that the settling velocity depends only on the densities of
particles via the parameter $\beta$.  Assuming a mean density of sea water in
the upper ocean as $\rho_f =1025 kg/m^3$ the parameter $\beta$ has values
within the range $[0.5,0.99]$ for the typical values of the density of marine
particles previously discussed. Fig. \ref{fig:modelvalidity} shows $v_s$ for
different sizes and the regions where $Re_p>1$ (and other parameter regions
where MRG is not a good approximation) as a function of particle radius and
for the limiting values of $\beta$. It reveals that Eq. (\ref{eq:MRG}) can not
describe ocean particles larger than $300\mu m$ of any density, and for a
limited range of densities when the particle radius exceeds approximately
$100 \mu m$. In fact, the range of application of MRG to marine particles is
plotted in the blue area, which at the same time gives an estimate of the
typical sinking velocities for a given particle size.

Summarizing, both the MRG and its approximation Eq.  (\ref{eq:MRsimp}) are
valid for marine particles with size within the range $1 \mu m$ and
$200 \mu m$. That is, it is valid for all particulate organic matter in Fig.
\ref{fig:sizeparticles} except the largest of the micro-aggregates and meso-
and macro-bodies of zooplankton. The sinking velocities range from $1 mm/day$-
$1 km/day$.

\subsection{The MRG equation in a rotating frame and further
simplications}

We are interested in applying Eq. (\ref{eq:MRsimp}) in oceanic flows, where
the particle $\vn$ and flow $\un$ velocities are expressed in a frame rotating
with the Earth angular velocity $\Omegan$
\citep{Elperin1998,Biferale2016,Tanga1996,Provenzale1999,Sapsis2009}. Both
time derivatives $\frac{d}{dt}$ and $\frac{D}{Dt}$ have to be corrected
following the rule
\begin{align}
  \frac{d}{dt} &\to \frac{d}{dt} + 2\Omega \times \vn + \Omegan \times (\Omegan \times \rn),\\
  \frac{D}{Dt} &\to \frac{D}{Dt} + 2\Omega \times \un + \Omegan \times (\Omegan \times \rn).
\end{align}
Where $\Omega=|\Omegan|$ and $\rn$ is the particle position vector whose
origin is in the rotation axis. So that Eq.(\ref{eq:MRsimp}) is now
\begin{equation}
  \frac{d\vn}{dt}=\beta \frac{D\un}{Dt} - \frac{\vn - \un}{\tau_p} - 2 \Omegan
  \times (\vn-\beta \un)  + \vn'_s/\tau_p.
\end{equation}

Two apparent forces arise in the equation, the Coriolis force
$2\Omegan\times(\vn-\beta\un)$ and the centrifugal force, which is included in
a modified sinking velocity
$\vn'_s=(1-\beta) (\gn-\Omegan \times (\Omegan \times \rn))
\tau_p$. The effect of the centrifugal force is very small (of order $10^{-3}$
compared to gravity) and can be absorbed in a redefinition of
$\mathbf{g}$. Thus, in the following we take $\vn'_s=\vn_s$ with the properly
chosen $\mathbf{g}$.

The ratio between the particle response time and the Kolmogorov time scale is
the Stokes number $St=\tau_p/\tau_{\eta}$, which measures the importance of
particle's inertia because of its size and density. According
  to the range of $\epsilon$ in the ocean mentioned before, we get
  $0.1~s < \tau_{\eta} < 5~s$, and for our range of particle sizes
  $10^{-6}~s < \tau_p < 10^{-2}~s$ so we can assume that $St<<1$ (see Fig.
  \ref{fig:modelvalidity}). This motivates us to make a second (standard)
approximation \citep{Balkovsky2001,Haller2008} of the MRG equation expanding
in powers of $\tau_p$ (note that it would be more natural to make the
expansion in powers of the non-dimensional $St$ but we prefer to do it in
$\tau_p$ to control on the time scales of the problem). Assuming first
the solution to Eq. (\ref{eq:MRsimp}):
\begin{equation*}
  \vn=\un + \un_1 \tau_p + \un_2 \tau^2_p + \dots,
\end{equation*}
and using $\frac{d\vn}{dt}=\frac{D\un}{Dt} + O(\tau_p)$, we get that the
particle velocity at first order in $\tau_p$ is
\begin{equation}\label{eq:final}
  \vn = \un + \vn_s + \tau_p (\beta-1) \left( \frac{D\un}{Dt} + 2 \Omegan \times \un \right).
\end{equation}

It is worth recalling that $\tau_p(1-\beta)=v_s/g$, so that all
dependencies on particle size and density appear in Eq.
(\ref{eq:final}) through the combination of parameters defining
$v_s$. Different combinations of size and density, taken within
the ranges reported in Sect.
\ref{sec:propertiesmarineparticle}, follow the same dynamics if
they have the same undisturbed settling velocity $v_s$.

A further discussion of Eq. (\ref{eq:final}) follows. At this order only three
physical processes correct the particle velocity with respect to the fluid
velocity: the Stokes friction determining the settling velocity $v_s$, the
inertial term given by $\tau_p (\beta-1) \frac{D\un}{Dt}$ whose major effect
is to introduce a centrifugal force pulling particles away from vortex cores
\citep{Maxey1987,Michaelides1997}, and the influence of the Coriolis force
$2\tau_p (\beta-1) \Omegan \times \un $. Concerning sinking dynamics, the
$\vn = \un + \vn_s$ is the most relevant approximation, and many other studies
consider it, mainly in oceanographic contexts \citep[e.g.][]{Siegel1997}.
Note that we can use the right-hand-side of Eq.  (\ref{eq:final}) with
$\un=\un(\rn,t)$ to define the particle velocity $\vn$ as a velocity field in
three-dimensional space $\vn=\vn(\rn,t)$. If one uses the lowest-order
approximation $\vn \approx \un$ we have
$\nabla \cdot \vn = \nabla \cdot \un =0$ when the fluid velocity field $\un$
is incompressible (which is the case for ocean flows). This means that when
considering this term alone, one cannot obtain a compressible particle
velocity whereas this was the main reason invoked to explain the clustering of
finite-size particles \citep{Squires91,Bec2003}. For this reason, numerous
studies \citep{Tanga1996,Michaelides1997,Bec2007,
  Bec2014,Cartwright2010,Guseva2013,Gustavsson2014,Beron-Vera2015} consider
the role of the additional terms. With them
$\nabla \cdot \vn = \tau_p (\beta - 1) \nabla \cdot ( \frac{D\un}{Dt}+ 2
\Omegan \times \un) \neq 0$, and inertia-induced clustering may
occur. In the following sections we address two main
questions: a) how relevant for the sinking dynamics are the Coriolis and
centrifugal terms?; and b) are they essentical ingredients for the clustering
of biogenic particles? We will study the relevance of the different terms in
Eq. (\ref{eq:final}) in a realistic oceanic setting.

\section{Numerical simulations}
\label{sec:simulations}

\begin{figure}[t]
\centering
\includegraphics[width=\columnwidth]{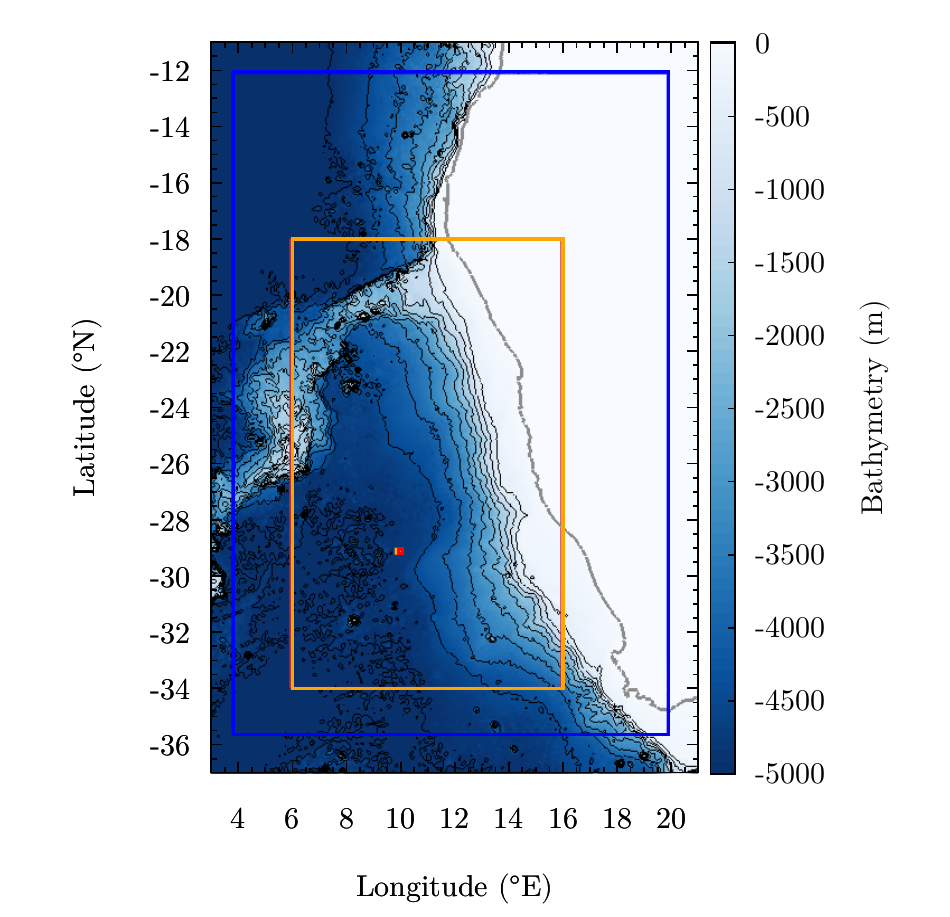}
\caption{Map of region of study. Color corresponds to bathymetry. Blue
  rectangle is region used for simulations of the ROMS model. Orange rectangle
  is the region for the clustering numerical experiment of
  Sect. \ref{sec:clustering} and red rectangle is the release site of the
  sinking numerical experiments of
  Sect. \ref{sec:simulations}. Gray represent the coastline.}
\label{fig:dataregion}
\end{figure}

The velocity flow $\un$ of the Benguela region was produced by a regional
simulation of a hydrostatic free-surface primitive equations model called ROMS
(Regional Ocean Modelling System). The configuration used here extends from
12°S to 35°S and from 4°E to 19°E (blue rectangle
in Fig. 3) and was forced with climatological atmospheric data
\citep{Gutnek2013}. The simulation area extends from $12^{\circ}S$ to
$35^{\circ}S$ and from $4^{\circ}E$ to $19^{\circ}E$ (blue rectangle in
Fig. \ref{fig:dataregion}). The velocity field data set consists of 2 years of
daily averages of zonal ($u$), meridional ($v$) and vertical velocity ($w$)
components, stored in a three-dimensional grid with a horizontal resolution of
$1/12^o$ and 32 vertical terrain-following levels using a stretched vertical
coordinate where the layer thickness increases from surface/bottom to the
ocean interior.

In order to integrate particle trajectories from the velocity in
Eq. (\ref{eq:final}) we interpolate linearly $\un(\rn,t)$ from the closest
space-time grid points to the actual particle locations. Given the huge
disparity between the model resolution and the small particle-sizes
considered, it is pertinent to parameterize in some way the unresolved
scales. This can be done by different approaches, from stochastic Lagrangian
modeling \citep{Brickman2002}, to deterministic kinematic fields
\citep{Palatella2014}. The first approach is adopted by adding a simple white
noise to the particle velocity \citep{Tang2012}, with different intensity in
the vertical and horizontal directions. Thus, we consider this noisy version
of the simplified MRG:
\begin{eqnarray}
\frac{d \rn(t)}{dt} &=& \vn(t)
\label{eq:MRGnoisyA} \\
\vn &=& \un + \vn_s + \tau_p (\beta-1) \left( \frac{D\un}{Dt}+ 2 \Omegan \times \un \right) + \Wn .
\label{eq:MRGnoisyB}
\end{eqnarray}
$\Wn(t) \equiv \sqrt{2D_h}\Wn_h(t) + \sqrt{2D_v}W_z(t)$, with
$(\Wn_h,W_z)=(W_x(t),W_y(t),W_z(t))$ a three-dimensional vector Gaussian white
noise with zero mean and correlations
$\langle W_i (t) W_j (t')\rangle=\delta_{ij} \delta (t-t')$, $i,j=x,y,z$. We
consider an horizontal eddy diffusivity, $D_h$, depending on resolution length
scale $l$ according to Okubo formula \citep{Okubo1971,Isma2011}:
$D_h(l)=2.055 \times 10^{4}l^{1.55}$ $(m^2/s)$.  Thus, if taking
$l \sim 8~km = 8000~m$ (corresponding to $1/12^{\circ}$) we obtain
$10m^2/s$. In the vertical direction we use a constant value of
$D_v=10^{-5}m^2/s$ \citep{Rossi2013}.


In order to obtain quantitative assessment of the relative effects of the
different physical terms in Eq.  (\ref{eq:MRGnoisyB}), we will compare
trajectories obtained from the following expressions which only consider some
of the terms of the full expression Eq. (\ref{eq:MRGnoisyB}):
\begin{align}
\vn^{(0)} &= \un + \vn_s + \Wn, \label{eq:3equationsa} \\
\vn^{(co)} &= \un + \vn_s + \tau_p (\beta-1) 2 \Omegan \times \un + \Wn, \label{eq:3equationsb} \\
\vn^{(in)} &= \un + \vn_s + \tau_p (\beta-1) \frac{D\un}{Dt} + \Wn .    \label{eq:3equationsc}
\end{align}
Besides the random noise term, the first expression (\ref{eq:3equationsa})
only considers the settling velocity, equation (\ref{eq:3equationsb}) resolves
the settling velocity plus the Coriolis effect, and equation
(\ref{eq:3equationsc}) considers the settling plus the inertial term.

For the numerical experiments we will consider a set of six
values of $v_s$ ranging from $5 m/day$ to $200 m/day$, with
different integration times to have in all the cases a sinking
to about $1000-1100~m$ depth. The stochastic equation
(\ref{eq:MRGnoisyA}) with expressions
(\ref{eq:MRGnoisyB})-(\ref{eq:3equationsc}) is written in
spherical coordinates and numerically integrated using a
second-order Heun's method with time step of 4 hours
\citep{Toral2014}.  We use $R=6371~km$ for the Earth radius,
$g=9.81 m/s^2$, and the angular velocity $\Omegan$ is a vector
pointing in the direction of Earth axis and modulus
$|\Omegan|=7.2722\times 10^{-5}~s^{-1}$. We
take $v_s$ and $\tau_p$ constant in each experiment because,
although water density may increase with depth, this variation
is at most of $10 kg/m^3$ in the range of depths we are
considering here and then the impact on $v_s$ is below 0.1\%.
We use as initial starting date 17 September 2008. The
numerical experiments consist in launching $N=6000$ particles
from initial conditions randomly chosen in a square of size
$1/6^{\circ}$ centered at $10.0^{\circ}E$ $29.12^{\circ}S$ and
$-100.0 m$ depth (red rectangle in Fig. \ref{fig:dataregion}),
 and in letting them evolve for a given time $t_f$
(stated in Table \ref{tab:meanfinaldepths}) following Eq.
(\ref{eq:MRGnoisyA}) with expressions
(\ref{eq:MRGnoisyB})-(\ref{eq:3equationsc}). We use in each
case identical initial conditions and the same sequence of
random numbers for the noise terms. In this way we guarantee
that any difference in particle trajectories arise from the
inclusion or not of the inertial and Coriolis terms. We obtain the
time-dependent positions of all the particles for each
approximation to the dynamics: $\rn_i(t)$, $\rn^{(0)}_i (t)$,
$\rn^{(co)}_i (t)$, and $\rn^{(in)}_i (t)$, $i=1,...,N$, following
respectively Eqs. (\ref{eq:MRGnoisyB})-(\ref{eq:3equationsc}) and
the corresponding final positions at $t=t_f$.

Table \ref{tab:meanfinaldepths} gives the mean and the standard deviation of
the depths attained by the set of particles in each numerical experiment as
obtained from Eqs.  (\ref{eq:MRGnoisyA}) and
(\ref{eq:MRGnoisyB}). We find that the use of the different
approximations (\ref{eq:3equationsa})-(\ref{eq:3equationsc}) gives virtually
the same results. The only differences larger than $1~cm$ in mean or standard
deviation are the ones for the smallest unperturbed settling velocity
considered, $v_s=5~m/day$, and are also reported in Table
\ref{tab:meanfinaldepths}. The measured differences are negligible as compared
with the traveled distance or even with the model grid size. Indeed small
changes in the ROMS model configuration or in the velocity interpolation
procedure would have an impact larger than this. The mean
  displacements in the horizontal obtained with the different approximations
  (not shown) differ also in less $0.1\%$. We thus conclude that the simplest
approximation Eq.  (\ref{eq:3equationsa}) which only considers passive
transport and an added constant sinking velocity already provides a good
description of the sinking process for the type of marine particles and the
range of space and time scales considered here. Note that the
  depth attained by the particles is always slightly shallower than
$z=-1100~m$, which is the depth that would be reached in a still fluid.  It is
still debated under which conditions fluid flows enhances or reduces the
settling velocity \citep{Maxey1987,Wang1993,Ruiz2004,Bec2014}.

\begin{table}[!h]
  \centering
  \begin{tabular}[width=.8\columnwidth]{|c|c|c|c|}\hline
    $v_s$  & integration time &Mean final depth & std final depth  \\
    ($m/day)$ & $t_f$ ($days$)        & ($m$)           &    ($m$) \\ \hline
    200 &   5 & -1091.78 & 3.88\\ \hline
    100 &  10 & -1065.33 & 6.57\\ \hline
     50 &  20 & -1033.97 & 6.22\\ \hline
     20 &  50 & -1051.85 & 22.67\\ \hline
     10 & 100 & -1043.49 & 51.22\\ \hline \hline
      5 & 200 & -1054.97 & 62.03\\
       &     &  -1054.76 \hfill(co)& 62.14 \hfill(co) \  \\
       &     &  -1054.76 \hfill(in)& 62.16 \hfill(in) \  \\
       &     &   -1054.72 \hfill(0)& 62.14 \hfill(0) \ \\
        \hline
  \end{tabular}
  \caption{Mean and standard deviation of the set of depths attained,
according to Eqs. (\ref{eq:MRGnoisyA}) and
(\ref{eq:MRGnoisyB}), by the set of particles released from the
red rectangle in Fig. \ref{fig:dataregion} at $z=-100~m$ for
the different values of $v_s$ and integration times used. The
results labeled $(co)$, $(in)$, and $(0)$ are obtained from the
different approximations in Eqs.
(\ref{eq:3equationsa})-(\ref{eq:3equationsc}), which differ
more than $1~cm$ from the ones obtained from Eq.
(\ref{eq:MRGnoisyB}) only in the $v_s=5~m/day$ case. }
\label{tab:meanfinaldepths}
\end{table}

We perform now a more stringent test going beyond the analyses of mean
displacements by considering differences between individual particle
trajectories. To assess the impact of the Coriolis and of the inertial effects
we compare the positions $\rn^{(co)}_i (t)$, and $\rn^{(in)}_i (t)$ with the
simpler dynamics Eq. (\ref{eq:3equationsa}) which gives $\rn^{(0)}_i (t)$ for
each time $t$. To do so we compute the root mean square difference in position
per particle and time, which we separate in vertical and horizontal
components:
\begin{eqnarray}
r_h^{(k)}(t) &=& \sqrt{\frac{1}{N}\sum_{i=1}^N
\left(\xn^{(0)}_i (t) - \xn^{(k)}_i (t)\right)^2}\\
r_v^{(k)}(t) &=& \sqrt{\frac{1}{N}\sum_{i=1}^N
\left(z^{(0)}_i (t) - z^{(k)}_i (t)\right)^2}
\end{eqnarray}
with $\xn_i=(x_i,y_i)$, the horizontal position vectors, and
the superindex $(k)$ takes the values $(co)$ or $(in)$.

%

\begin{figure}[!h]
\includegraphics[width=\columnwidth]{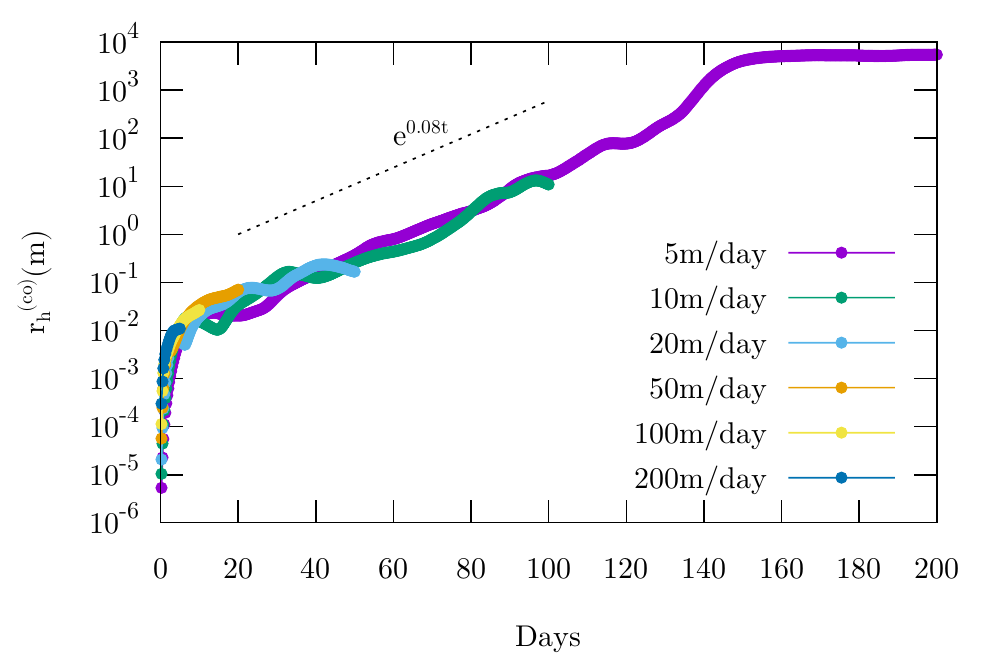}
\caption{
Root mean square difference per particle, as a function of time,
between horizontal particle positions computed with Eq. (\ref{eq:3equationsa}) and with
Eq. (\ref{eq:3equationsb}), i.e. with and without the Coriolis term. The different colors correspond to
distinct values of the unperturbed sinking velocity. The dashed line is an exponential with
slope $0.08~day^{-1}$.
}
\label{fig:hdispcoriolis}
\end{figure}

Fig. \ref{fig:hdispcoriolis} shows the influence of the Coriolis term in the
horizontal component for each sinking velocity as a function of time. We
observe an exponential growth in a wide range of times, which reveals the
chaotic behavior of each of the compared trajectories.  The value of the
exponent $~0.08 days^{-1}$ is in agreement with the order of magnitude of the
Lyapunov exponent calculated using the same ROMS velocity model and region
\citep{Bettencourt2012}. Similar exponential growth with the same growth rate
were observed for the inertial terms and the vertical components (not shown),
although the absolute magnitude of these mean root square differences was much
smaller.

\begin{figure}[!h]
\includegraphics[width=\columnwidth]{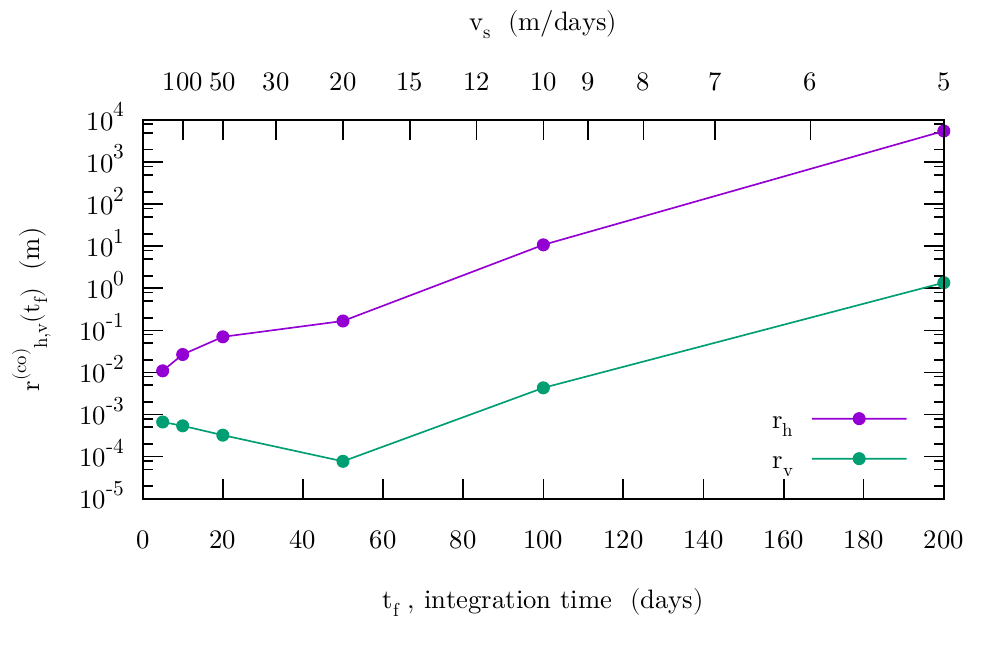}
\caption{Root mean square difference per particle between final positions
(at times $t_f$ stated in Table \ref{tab:meanfinaldepths}) computed with and without the
Coriolis term (Eqs. (\ref{eq:3equationsb}) and (\ref{eq:3equationsa}), respectively).
Data are presented as a function of the unperturbed sinking velocity $v_s$ used
(upper horizontal scale) and of the final integration time $t_f$ (lower horizontal scale). Upper violet line,
the horizontal difference $r_h^{(co)}(t_f)$; lower green line, the vertical difference $r_v^{(co)}(t_f)$. }
\label{fig:coriolis}
\end{figure}

\begin{figure}[!h]
\includegraphics[width=\columnwidth]{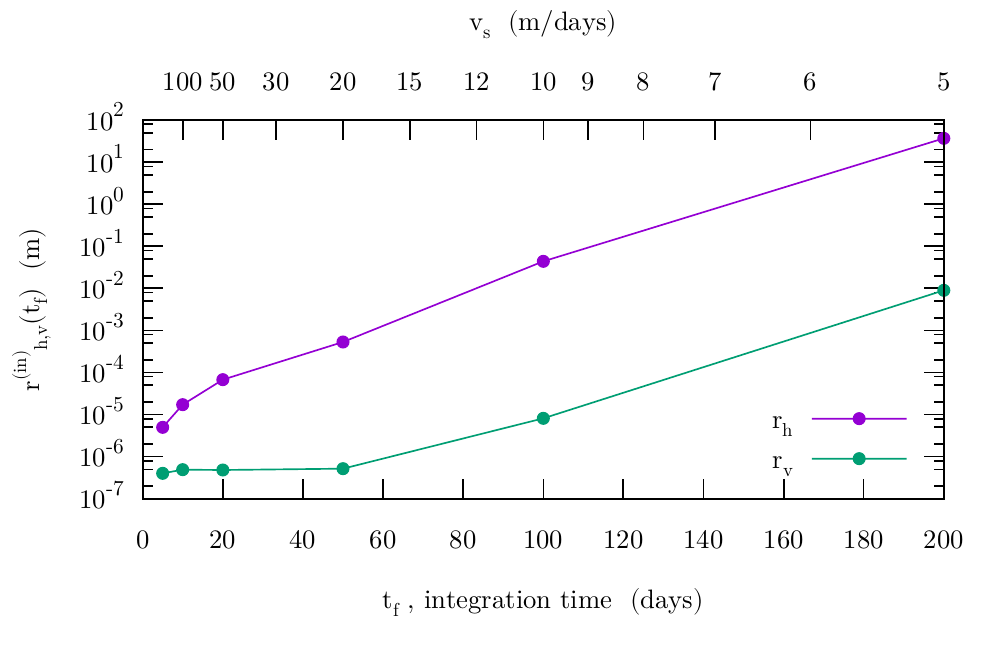}
\caption{Root mean square difference per particle between final positions
(at times $t_f$ stated in Table \ref{tab:meanfinaldepths}) computed with and without the
inertial term (Eqs. (\ref{eq:3equationsc}) and (\ref{eq:3equationsa}), respectively).
Data are presented as a function of the unperturbed sinking velocity $v_s$ used
(upper horizontal scale) and of the final integration time $t_f$ (lower horizontal scale). Upper violet line,
the horizontal difference $r_h^{(in)}(t_f)$; lower green line, the vertical difference $r_v^{(in)}(t_f)$. }
\label{fig:centrifugal}
\end{figure}

The horizontal and vertical differences $r_{h,v}^{(co)}$ at the final
integration time $t_f$ (i.e. the time at which the particles reach an
approximate depth of $1000~m$ for each value of $v_s$) are displayed in Fig.
\ref{fig:coriolis}, both as a function of $v_s$ and of $t_f$.  Similarly, the
values of $r_{h,v}^{(co)}$ are presented in Fig.  \ref{fig:centrifugal}. The
behavior can be understood as resulting from two factors: on the one hand
smaller $v_s$ requires larger $t_f$ to reach the final depth, and larger
integration time $t_f$ allows for accumulation of larger differences between
trajectories. On the other hand the Coriolis and inertial terms in Eqs.
(\ref{eq:3equationsb})-(\ref{eq:3equationsc}) are proportional to
$\tau_p(\beta-1)=v_s/g$ so that their magnitude decreases for smaller
$v_s$. The combination of these two competing effects shapes the curves in
Figs. \ref{fig:coriolis} and \ref{fig:centrifugal}, which for the
vertical-difference case turn-out to be non-monotonic in $v_s$ or $t_f$.

In all cases, the differences (both in vertical and horizontal) between the
simple dynamics (\ref{eq:3equationsa}) and the corrected ones in
Eqs. (\ref{eq:3equationsb}) and (\ref{eq:3equationsc}) are negligible when
compared with typical particle displacements, or even with model grid sizes.
For example, we imposed in our simulations a vertical displacement close to
1000~m, whereas the mean root square difference with respect to simple sinking
is below $1~m$ for the Coriolis case (Fig. \ref{fig:coriolis}) and below
$1~cm$ for the inertial case (Fig. \ref{fig:centrifugal}). In the horizontal
direction, displacements during those times are of the order of hundreds of
$km$, whereas the corrections introduced by the Coriolis and inertial terms
are in the worst cases of the order of a few kilometers or of tens of meters,
respectively. In particular, the most important impact (horizontal differences
of tens of kilometer) is attributed to the Coriolis term for particles sinking
at 5 m/day (Fig. \ref{fig:coriolis}). This indicates that the inclusion of the
Coriolis term would be required to properly model slowly sinking particles at
high latitudes. It is worth noting that although the small value of Rossby
number $\simeq 0.01$ for mesoscale processes might indicate a strong influence
of the Coriolis force in Eq. (\ref{eq:MRGnoisyB}), its influence on particle
dynamics becomes negligible because it is multiplied by $\tau_p$ or
equivalently, the Stokes number, which is significantly small for biogenic
particles. Nevertheless Rossby number coincides with the ratio of inertial
term to Coriolis term in Eq. (\ref{eq:MRGnoisyB}) and its value $\simeq 0.01$
explains the difference of two orders of magnitude among the corrections
arising from the inertial force and from Coriolis. The trajectories of the
full dynamics ruled by Eq.  (\ref{eq:MRGnoisyB}) are nearly identical to the
ones under the approximation which keeps only the sinking term and Coriolis,
so that the corresponding comparison to $\rn_i^{(0)}$ gives a figure
essentially identical to Fig. \ref{fig:coriolis} (not shown).

In summary, for the range of sizes and densities of the marine
particles considered here, the sinking dynamics is essentially
given by the velocity $\vn = \un + \vn_s$, which has been the
one used in some oceanographic studies
\citep{Siegel1997,Siegel2008,Roullier2014}. Note however that a
new question arises: what is then the reason for the observed
clustering of falling particles
\citep{Logan1990,Buesseler2007,Mitchell2008}?  The argument of
the non-inertial dynamics of the particles does not serve since
$\nabla \cdot \vn = \nabla \cdot \un = 0$. A possible response
is explored in the next section.

\section{Geometric clustering of particles}
\label{sec:clustering}

Compressibility of the particle-velocity field, i.e.
$\nabla\cdot \vn \neq 0$, which can arise from inertial effects even when the
corresponding fluid-velocity field is incompressible $\nabla\cdot \un = 0$,
has been identified as one of the mechanisms leading to preferential
clustering of particles in flows \citep{Squires91,Balkovsky2001}. This is so
because $\rho(t)$, the particle density at time $t$ at the location
$\rn=\rn(\rn_0,t)$ of a particle that started at $\rn_0$ at time zero,
satisfies $\rho(t)=\rho(0) \delta^{-1}$, where $\delta$ is a dilation factor
equal to the determinant of the Jacobian
$|\frac{\partial\rn}{\partial \rn_0}|$, which satisfies
\begin{equation}
\frac{1}{\delta}\frac{D\delta}{Dt}=\nabla \cdot \vn
\end{equation}
or, using $\delta(0)=1$:
\begin{equation}
\delta (t_f)=e^{\int_0^{t_f}  dt \nabla \cdot \vn } \ .
\end{equation}
Thus, particles will accumulate (i.e. higher $\rho(t_f)$) in final deep
locations receiving particles whose trajectories have predominantly travelled
through regions with $\nabla\cdot \vn < 0$. We have seen however that, to a
good approximation $\nabla\cdot \vn \approx \nabla\cdot \un =0$ since inertial
effects can be neglected for the type of marine particles we consider here,
and then the three-dimensional particle-velocity field is incompressible.

\begin{figure*}[!h]
\includegraphics[width=\textwidth]{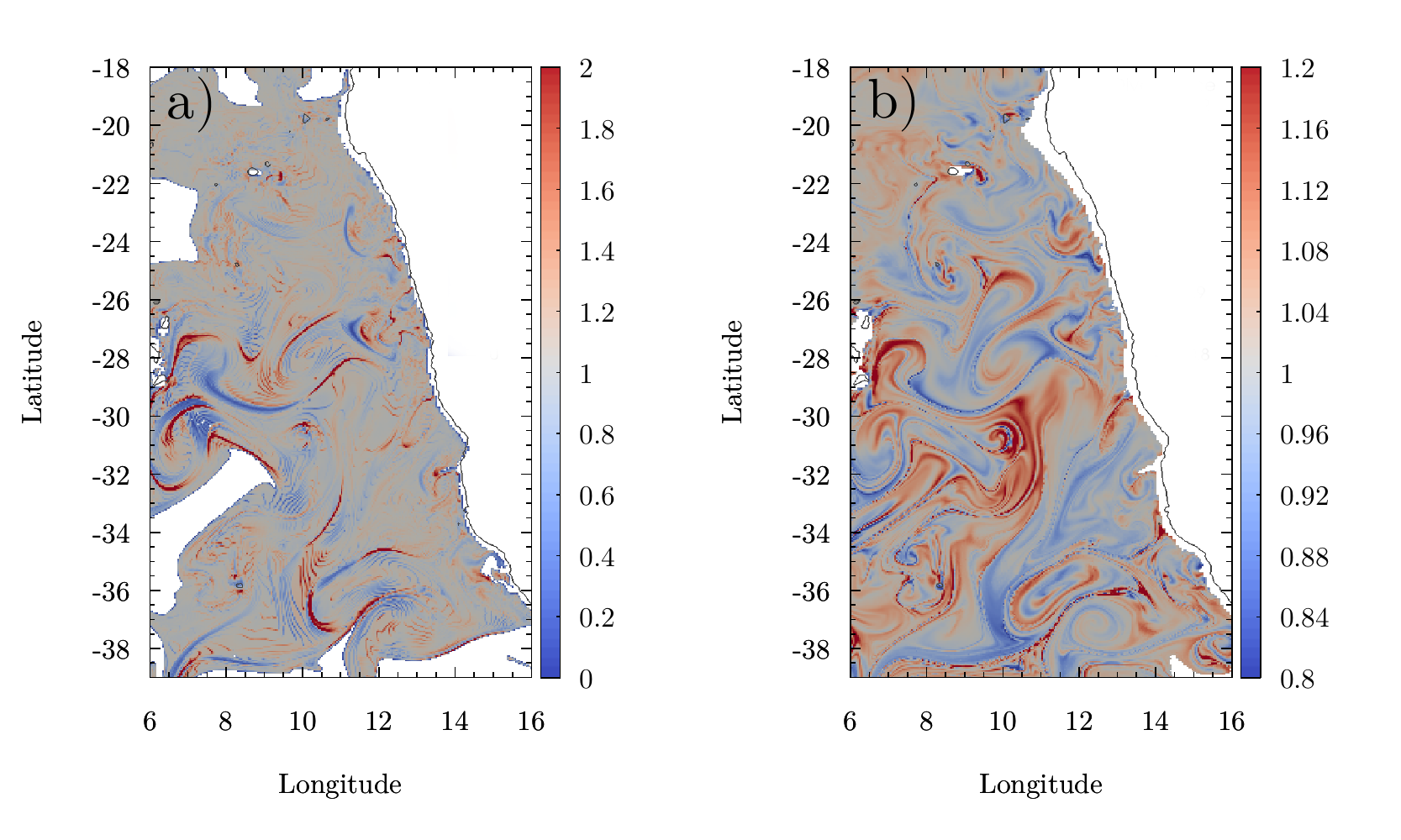}
\caption{ Results of the clustering numerical experiments of
  Sect. \ref{sec:clustering}. a) $N_f/N_0$, the number of particles $N_f$
  arriving to an horizontal box of size $1/25^{\circ}$ in the horizontal layer
  at $z=-1000~m$, normalized by the number of particles $N_0=1000$ released
  from the upper $z=-100~m$ layer. b) The corrected dilation factor
  $\delta(\xn,t_f)^{-1} \cos(\theta_f)/\cos(\theta_0)$ mapped on the final
  $z=-1000~m$ layer. It gives the ratio between horizontal densities at the
  final and initial locations, corrected with the latitudinal dependence of
  the horizontal boxes used in panel a), to give an estimation of the local
  particle number ratio between lower and upper layer. Black thin line
  represents the coastline; white oceanic areas indicate in a) regions which
  do not receive any falling particles; in b) regions from which the backward
  integration ends up outside the domain.}
\label{fig:clustering}
\end{figure*}

We now reproduce numerically a typical situation in which clustering of marine
particles is observed. We release particles uniformly in an horizontal layer
close to the surface, we let them sink within the oceanic flow and we finally
observe the distribution of the locations where they touch another horizontal
deeper layer. The domain chosen is the rectangle $12^{\circ}S$ to
$35^{\circ}S$ and $4^{\circ}E$ to $19^{\circ}E$ (orange rectangle in
Fig. \ref{fig:dataregion}).  We divide the domain horizontally in squares of
side $1/25^{\circ}$, then initialize 1000 particles at random positions in
each of them in August 20, 2008 at depth $z=-100~m$ (i.e. the bottom of the
euphotic layer, starting point of our biogenic particles), and then integrate
each trajectory until it reaches $-1000m$ depth. We use expression
(\ref{eq:3equationsa}) for the velocity, with $v_s=50m/day$. In order to avoid
any small fluctuating compressibility arising from the noise term we put
$\Wn=\mathbf{0}$ but we have checked that the result in the presence of noise
is virtually indistinguishable (not shown). At the bottom layer ($z=-1000~m$)
we count how many particles arrive to each of the $1/25^{\circ}$ boxes and
display the result in Fig.  \ref{fig:clustering}(a). Despite
$\nabla\cdot \vn =0$ we see clear preferential clustering of particles in some
regions related to eddies and filaments. We note that our horizontal boxes
have a latitude-dependent area so that distributing particles at random in
them produces a latitude-dependent initial density which could lead to some
final inhomogeneities. We have checked however that for the
range of displacements of the particles, this effect is everywhere smaller
than $5\%$ and thus can not be responsible for the large clustering observed
in Fig. \ref{fig:clustering}(a). Nevertheless, this effect will be taken into
account later.

We explain the observed particle clustering by considering the field displayed
in Fig. \ref{fig:clustering}(a) as a projection in two dimensions of a density
field (the cloud of sinking particles) which evolves in three-dimensions. Even
if the three-dimensional divergence is zero, and then an homogeneous
three-dimensional density will remain homogeneous, a two-dimensional cut or
projection can be strongly inhomogeneous. This mechanism has been proposed to
explain clustering and inhomogeneities in the ocean surface
\citep{Huntley2015,Jacobs2016}, but we show here that it is also relevant for
the crossing of a horizontal layer by a set of falling particles.

A simple way to confirm that this clustering arises from the
two-dimensionality of the measurement is to estimate the changes in the
horizontal density of evolving particle layers as if they were produced just
by the horizontal part of the velocity field. This is only correct if an
initially horizontal particle layer remains always horizontal during the
sinking process, which is not true. But, given the huge differences in the
values of the horizontal and vertical velocities in the ocean, we expect this
approximation to capture the essential physics and provide a qualitative
explanation of the observed clustering. We expect the approximation to become
better for increasing $v_s$, because of the shorter sinking time during which
vertical deformations could develop. Thus we compute the two-dimensional
version of the dilation field, $\delta_h(\xn,t_f)$, at each horizontal
location $\xn$ in the deep layer at $z=-1000m$:
\begin{equation}\label{eq:deltah}
\delta_h (\xn,t_f) = e^{\int_0^{t_f} dt \nabla_h\cdot \vn}
\end{equation}
with the horizontal divergence
\begin{equation}
\nabla_h \cdot \vn\equiv \frac{\partial v_x}{\partial x}+
\frac{\partial v_y}{\partial y}=\frac{\partial u}{\partial x}+
\frac{\partial v}{\partial y}=-\frac{\partial w}{\partial z} \ ,
\end{equation}
where in the second equality we have used Eq.  (\ref{eq:3equationsa}) from
which $\nabla_h \cdot \vn= \nabla_h \cdot \un$ and the third one is a
consequence of $\nabla\cdot \un = 0$.  In order to get the values of
$\delta_h$ on a uniform grid on the $-1000m$ depth layer at the arrival date
$t_f$ of the particles in the previous simulation, we integrate backwards in
time trajectories from grid points separated $1/50^{\circ}$ at $z=-1000m$
until they reach $-100m$.  The starting date ($t_f$) of the backwards
integration was September 7, 2008, i.e. 18 days after the release date used in
the previous clustering experiment.  This value correspond to the average
duration time of trajectories in that experiment. Then $\delta_h$ was computed
integrating in time the values of $\nabla_h\cdot\vn$ along every trajectory
using Eq.  (\ref{eq:deltah}).

Figure \ref{fig:clustering}(b) displays the quantity
$\delta(\xn,t_f)^{-1} \cos(\theta_f)/\cos(\theta_0)$, which gives the ratio
between densities in the upper and lower layer, corrected with the angular
factors controlling the area of the horizontal boxes so that this can be
compared with the ratio between particle numbers displayed in Fig.
\ref{fig:clustering}(a). $\theta_f$ is the latitude of point $\xn$, and
$\theta_0$ is the latitude of the corresponding trajectory in the upper
$z=-100~m$ layer.  As stated before, the latitudinal corrections by the cosine
terms are always smaller than a $5\%$.  Although there is no perfect
quantitative agreement, there is clear correspondence between the main
clustered structures in panels (a) and (b) of Fig. \ref{fig:clustering},
confirming that they originate from the horizontal dynamics in an
incompressible three-dimensional velocity field.  We have checked in specific
cases that locations with larger differences between Figs.
\ref{fig:clustering}(a) and (b) correspond to places with large dispersion in
the arrival times to the bottom layer, indicating deviations from the
horizontality assumption.

\conclusions  

We have studied the problem of sinking particles in a realistic oceanic flow,
focussing in the range of sizes and densities appropriate for marine biogenic
particles. Starting from a modeling approach in terms of the MRG equation
(\ref{eq:MRG}), our conclusion is that the simplest approximation given by
Eq. (\ref{eq:3equationsa}) in which particles move passively with the fluid
flow with an added constant settling velocity in the vertical direction is an
accurate framework to describe the sinking process in the type of flows and
particles considered. A re-assessment of these assumptions may be required if
more complex processes (such as aggregation/disaggregation) are included and
when super-high resolution (submesoscale and below) mimicking the real ocean
will become available.

Corrections arising from the Coriolis force turn out to be about 100 times
larger than the ones coming from inertial effects, in agreement with the
results in \citet{Sapsis2009} or in \citet{Beron-Vera2015}, but both of them
are negligible when compared to the effects of passive transport by the fluid
velocity plus the added gravity term, except for very slowly sinking particles
in high latitudes.

If the fluid flow field $\un(\rn,t)$ has vanishing divergence then the same is
true for the particle velocity field defined by the approximation in
Eq. (\ref{eq:3equationsa}). Then, no three-dimensional clustering can occur
within this approximation. Nevertheless, we have shown that two-dimensional
cuts or projections of evolving three-dimensional particle clouds display
horizontal clustering.


\section{Data availability}
Data results are available in https://doi.org/10.20350/digitalCSIC/8504 \cite{Monroy2017}

%

%


\competinginterests{The authors declare that they have no
conflict of interest.}


\begin{acknowledgements}
  We acknowledge support from the LAOP project,
CTM2015-66407-P (AEI/FEDER, EU), from the Office of Naval
Research Grant No. N00014-16-1-2492, and through a Juan de la
Cierva Incorporaci\'{o}n fellowship (IJCI-2014-22343) granted
to V.R.
\end{acknowledgements}




%



\bibliographystyle{copernicus}
\bibliography{modelbiogenicparticles}

\end{document}